\documentclass[amsmath,amssymb,showkeys]{revtex4}
\usepackage{graphicx}

\usepackage{pstricks}
\usepackage{pst-node}
\usepackage{pst-plot}

\input{Qcircuit.tex}

\begin{document}
\title{Quantum Booth's array Multiplier}

\author{J.J. \'Alvarez-S\'anchez\footnote{Work done partially at the
Laboratoire d'Informatique Th\'eorique et Quantique, University of
Montreal, Canada} and J.V. \'Alvarez-Bravo}
\affiliation{Departamento de Inform\'atica, E.U. de Inform\'atica,
Universidad de Valladolid, 40005 Segovia, Spain}
\author{L.M. Nieto}
\affiliation{Departamento de F\'{\i}sica Te\'orica, At\'omica y
\'Optica, Facultad de Ciencias, Universidad de Valladolid, 47071
Valladolid, Spain}

\begin{abstract}
A new quantum architecture for multiplying signed integers is
presented based on Booth's algorithm, which is well known in
classical computation. It is shown how a quantum binary chain
might be encoded by its flank changes, giving the final product in
2's-complement representation.
\end{abstract}

 \vspace*{3pt} 

\maketitle

\section{Introduction and motivation}

After some spectacular results
\cite{Shor1,Grover1,BennettBrassard1}, people faced with the
fundamental issue of how could they build quantum architectures
for working out elementary arithmetic calculus \cite{VBE,
ripple-carry-add, logarithmic-carry-add, draper-add}. Building a
quantum computer implies to design a Quantum Arithmetic Logic Unit
and addition is the most fundamental arithmetic operation. Thus,
it has been the target for quantum arithmetic researchers that may
be divided in two groups:
\begin{enumerate}
\item Those that make the quantization of classical arithmetic
algorithms \cite{VBE,ripple-carry-add,logarithmic-carry-add}.

\item Those that use the Quantum Fourier Transform for building
adders \cite{draper-add}.
\end{enumerate}
Our point of view for making the quantization of the
multiplication operation is the first one. In
this work, we want to go a step further and tackle with the
multiplication of signed integers, which is not a trivial
arithmetic operation, even under the classical computation
paradigm. Multiplying signed integers needs a new information
representation in order to carry it out in a proper and efficient
way. Furthermore, multiplication involves addition of partial
products and bit scrollings. It is well known that signed integers
multiplication is usually performed encoding the multiplier and
performing additions, subtractions and bit displacements. This
algorithm is called Booth's algorithm \cite{Stallings}. The
mainidea of this algorithm is representing binary data as unsigned
integers for purposes of addition. This can be achieved by the
2's-complement representation, that makes possible performing
subtraction by means of addition~\cite{Mano}. With that in mind we
can work out additions and subtractions in a very easy way and
tackle the signed integers multiplication.

In this paper a Quantum Booth Multiplier (QBM) is presented {\em
based} on the corresponding Classical Booth's Algorithm, which is
described below.

\subsection{Classical Booth's algorithm}

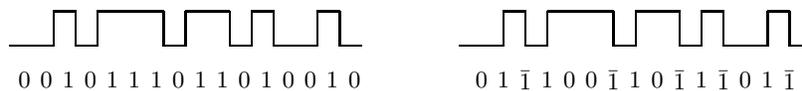
\begin{figure} [htbp]
\vspace*{-0.4cm}
\begin{center}
\begin{align*}
\Qcircuit @C=.9em @R=1.4em {
&& & \qw & & \qw & \qw & \qw & & \qw & \qw & & \qw & & & \qw &\\
&\qw & \qwx\qw & \qwx & \qwx\qw & & & \qwx & \qwx\qw & & \qwx &
\qwx\qw & \qwx & \qw & \qwx\qw & \qwx & \qw\\
\hspace{.4cm}0&\hspace{.4cm}0&\hspace{.4cm}1&\hspace{.4cm}0&\hspace{.4cm}1&\hspace{.4cm}1&\hspace{.4cm}1&\hspace{.4cm}0
&\hspace{.4cm}1&\hspace{.4cm}1&\hspace{.4cm}0&\hspace{.4cm}1&\hspace{.4cm}0&\hspace{.4cm}0&\hspace{.4cm}1&\hspace{.4cm}0&\\
} \qquad\qquad \Qcircuit @C=.9em @R=1.4em {
&& & \qw & & \qw & \qw & \qw & & \qw & \qw & & \qw & & & \qw &\\
&\qw & \qwx\qw & \qwx & \qwx\qw & & & \qwx & \qwx\qw & & \qwx & \qwx\qw & \qwx & \qw & \qwx\qw & \qwx & \qw\\
&0&1&\bar{1}&1&0&0&\bar{1}&1&0&\bar{1}&1&\bar{1}&0&1&\bar{1}\\
}
\end{align*}
\end{center}
\caption{On the right part of
the figure is Booth's encoding of the 16 bits chain shown on the left.}\label{cronograma}
\end{figure}

\begin{figure} [htbp]
\includegraphics[width=11.6cm]{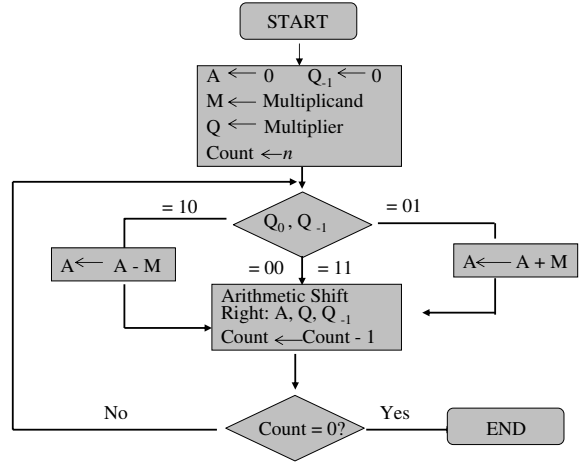}
\caption{\label{organigrama}Flow chart for
classical Booth's algorithm.}
\end{figure}

The Classical Booth's Algorithm encodes binary chains by means of
their transitions between 0's and 1's as it is shown in
Fig.~\ref{cronograma}. Once these transitions are detected, the
necessary displacements and additions are performed. This scheme
extends to any number of blocks of 1's in a multiplier, including
the case in which a single 1 is treated as a block. Indeed, we
might say that Classical Booth's Algorithm detects 1's
``islands''.
It is well known that multiplication can be achieved by adding
appropriately shifted copies of the multiplicand and that the
number of such operations can be reduced to two if we observe that
\begin{equation}
2^n\,+\,2^{n-1}\,+\,\cdots\,+\,2^{n-k}\,=\,2^{n+1}\,-\,2^{n-k} .
\end{equation}
\begin{figure}[!h]
\begin{center}
\includegraphics[width=9cm]{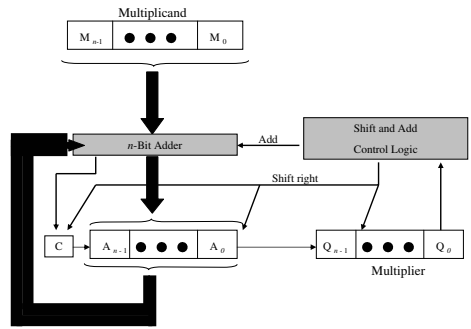}
\caption{Classical multiplier architecture}\label{arquitectura}
\end{center}
\end{figure}
Therefore, we encode the binary alphabet $\{1,0\}$ in another one,
$\{0,1, -1\}$, that shows the transitions among the binary logic
states within a binary chain. Thus, when a $-1$ is faced a
subtraction and a shift are performed (using the 2's-complement
representation) and when a $1$ is faced an addition and a shift
are performed. In the other two cases just an arithmetic shift is
performed for each situation. The algorithm conforms to this
scheme by performing a substraction when the first 1 of the block
is encountered ($1 \to 0$) and an addition when the end of the
block is encountered ($0 \to 1$). It is straightforward to show
that the same scheme works for a negative multiplier using the
2's-complement notation. The architecture of Booth's algorithm is
shown in Fig.~\ref{arquitectura},
and an example for the multiplication of 7 (0111) by 3 (0011), is
shown in Fig.~\ref{ejemploclasico}. The multiplier and the
multiplicand are placed in the $Q$ and $M$ registers,
respectively. There is also a 1-bit register placed logically to
the right of the less significant bit ($Q_0$) of the $Q$ register
and designated $Q_{-1}$. The results of the multiplication will
appear in the $A$ and $Q$ registers, the first one initialized to
0.
\begin{figure} [htbp]
\vspace*{-0.3cm}
\begin{center}
$$
\boxed{
\begin{array}{ccccl}
A & Q &Q_{-1} & M \\
0000 & 0011 & 0 & 0111 & {\rm Initial\ Values}
\\ [3ex]
\begin{array}{c} 1001 \\ 1100 \end{array} & \begin{array}{c} 0011 \\
1001 \end{array} &
\begin{array}{c} 0 \\ 1 \end{array} & \begin{array}{c} 0111 \\ 0111
\end{array} &
\left. \begin{array}{l} A\quad A-M \\ {\rm Shift} \end {array}
\right\}
\begin{array}{l} {\rm First} \\ {\rm Cycle} \end{array}
\\ [3ex]
\begin{array}{c} 1110 \end{array} & \begin{array}{c} 0100 \end
{array} &
\begin{array}{c} 1 \end{array} & \begin{array}{c} 0111 \end{array} &
\left. \begin{array}{l} {\rm Shift} \phantom{xxxxx}\end {array}
\right\}
\begin{array}{l} {\rm Second} \\ {\rm Cycle} \end
{array}
\\ [3ex]
\begin{array}{c} 0101 \\ 0010 \end{array} & \begin{array}{c} 0100 \\
1010 \end{array} &
\begin{array}{c} 1 \\ 0 \end{array} & \begin{array}{c} 0111 \\ 0111
\end{array} &
\left. \begin{array}{l} A\quad A+M \\ {\rm Shift} \end {array}
\right\}
\begin{array}{l} {\rm Third} \\ {\rm Cycle} \end{array}
\\ [3ex]
\begin{array}{c} 0001 \end{array} & \begin{array}{c} 0101 \end
{array} &
\begin{array}{c} 0 \end{array} & \begin{array}{c} 0111 \end{array} &
\left. \begin{array}{l} {\rm Shift} \phantom{xxxxx}\end {array}
\right\}
\begin{array}{l} {\rm Fourth} \\ {\rm Cycle} \end{array}
\end{array}
}
$$
\end{center}
\caption{\label{ejemploclasico}Example of Booth's algorithm.}
\end{figure}
Control logic scans the bits of $Q_{0}$ and $Q_{-1}$ one at a time.
According to Fig.~\ref{organigrama},
if the two bits are the same,
then all the bits of $A$, $Q$, and $Q_{-1}$ are shifted to
the right one bit. If the two bits differ, then the multiplicand
is added or subtracted from the $A$ register, according as the two
bits are $(0\to 1)$ or $(1\to 0)$. Following the addition or
subtraction, the right shift occurs. In either case, the right
shift is such that the leftmost bit of $A$, namely $A_{n-1}$, not
only is shifted into $A_{n-2}$, but also remains in $A_{n-1}$.
This is required to preserve the sign of the number in $A$ and $Q$
and it is named \emph{arithmetic shift}, since it preserves the
sign bit.

\section{Quantum Booth Multiplier}

\begin{figure} [htbp]
\begin{center}
\begin{minipage}[c]{9cm}
\Qcircuit @C=1.1em @R=1.4em {
  \lstick{\ket{x_{i+1}}_{1}} & \ctrl{2} & \ctrl{2}& \qw & \rstick{\ket{x_{i+1}}_{1}} \\
   \lstick{\ket{0}_{2}}& \qw & \qswap & \qw & \rstick{\ket{x'_{i+1,i}}_{2}} \\
  \lstick{\ket{x_{i}}_{3}}& \targ & \qswap & \qw & \rstick{\ket{x''_{i+1,i}}_{3}}
}
\end{minipage}
\end{center}
\caption{\label{rep_dos_qubits}Booth's encoding of
two adjacent qubits.}
\end{figure}
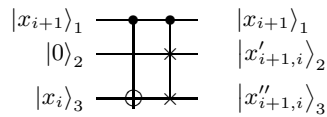

In this section a QBM is presented. First of all, we have to
encode the computational basis states $\{|0\rangle, |1\rangle\}$
that generate the Hilbert's space $\mathcal{H}^2$ by the new
alphabet $\{0,1, -1\}$ that will be represented by three
orthonormal quantum states belonging to the computational basis
that generates $\mathcal{H}^2\bigotimes\mathcal{H}^2$. These three
states are $|00\rangle\,\equiv\,0$, $|01\rangle\,\equiv\,1$, $|10
\rangle\,\equiv\,-1$, and have been chosen in this way because
they form a Gray code of three elements: all of them just differ
from each other in one binary digit. This will be a useful
property that will simplify the control stage in the QBM
architecture. Thus, the encoder, transforms the states couple
$\ket{x_{i+1}}_{1}$$\ket{x_{i}}_{3}$ in its transformed
$\ket{x'_{i+1,i}}_{2}$$\ket{x''_{i+1,i}}_{3}$ by means of the
table given in Fig.~\ref{transformaciones}, representing the
quantum operations performed by the quantum circuit that appears
in Fig.~\ref {rep_dos_qubits}.

\begin{figure} [htbp]
\begin{center}
\begin{minipage}[c]{9cm}
\Qcircuit @C=4.1em @R=1.4em {
&&& CNOT\hspace{1.8cm}&&CSWAP\hspace{1.8cm}&\\
&&\lstick{\ket{0_{1},0_{2},0_{3}}} & \longrightarrow
\hspace{1.8cm} &\lstick{\ket{0_{1},0_{2},0_{3}}} & \longrightarrow
\hspace{1.8cm}&
\lstick{\ket{0_{1},0_{2},0_{3}}} & \\
&&\lstick{\ket{0_{1},0_{2},1_{3}}} & \longrightarrow
\hspace{1.8cm} &\lstick{\ket{0_{1},0_{2},1_{3}}} & \longrightarrow
\hspace{1.8cm}&
\lstick{\ket{0_{1},0_{2},1_{3}}} & \\
&&\lstick{\ket{1_{1},0_{2},0_{3}}} & \longrightarrow
\hspace{1.8cm} &\lstick{\ket{1_{1},0_{2},1_{3}}} & \longrightarrow
\hspace{1.8cm}&
\lstick{\ket{1_{1},1_{2},0_{3}}} & \\
&&\lstick{\ket{1_{1},0_{2},1_{3}}} & \longrightarrow
\hspace{1.8cm} &\lstick{\ket{1_{1},0_{2},0_{3}}} & \longrightarrow
\hspace{1.8cm}&
\lstick{\ket{1_{1},0_{2},0_{3}}} & \\
}
\end{minipage}
\end{center}
\caption{\label{transformaciones}Transforming the
state of two adjacent qubits.}
\end{figure}

\begin{figure} [htbp]
\begin{center}
\begin{minipage}[c]{9cm}
{
{ \Qcircuit @C=.8em @R=.9em {
&&&&&&&&&&&&&&&&&&&\\
&&& \lstick{\ket{x_{3}}_{1}} & \qw & \qw      & \ctrl{4} & \ctrl{4}& \qw & \qw & \qw & \rstick{\ket{x_{3}}_{1}} \\
&&&&&&&&&&&&&&&&&&&\\
&&& \lstick{\ket{0}_{2}}     & \qw & \qw      & \qw      & \qswap  & \qw & \qw & \qw & \qw & \qw & \qw & \qw & \qw & \rstick{\ket{x'_{3,2}}_{2}}\\
&&&&&&&&&&&&&&&&&&&\\
&&& \lstick{\ket{x_{2}}_{3}} & \qw & \ctrl{2} & \targ    & \qswap  & \qw & \qw & \qw & \qw & \qw & \qw & \qw & \qw &  \rstick{\ket{x''_{3,2}}_{3}}\\
&&&&&&&&&&&&&&&&&&&\\
&&& \lstick{\ket{0}_{1}}     & \qw & \targ    & \ctrl{3} & \ctrl{3}& \qw & \qw & \qw & \rstick{\ket{x_{2}}_{1}} \\
&&& \lstick{\ket{0}_{2}}     & \qw & \qw      & \qw      & \qswap  & \qw & \qw & \qw & \qw & \qw & \qw & \qw & \qw & \rstick{\ket{x'_{2,1}}_{2}}\\
&&&&&&&&&&&&&&&&&&&\\
&&& \lstick{\ket{x_{1}}_{3}} & \qw & \ctrl{2} & \targ    & \qswap  & \qw & \qw & \qw & \qw & \qw & \qw & \qw & \qw & \rstick{\ket{x''_{2,1}}_{3}}\\
&&&&&&&&&&&&&&&&&&&\\
&&& \lstick{\ket{0}_{1}}     & \qw & \targ    & \ctrl{3} & \ctrl{3}& \qw & \qw & \qw & \rstick{\ket{x_{1}}_{1}}\\
&&& \lstick{\ket{0}_{2}}     & \qw & \qw      & \qw      & \qswap  & \qw & \qw & \qw & \qw & \qw & \qw & \qw & \qw & \rstick{\ket{x'_{1,0}}_{2}}\\
&&&&&&&&&&&&&&&&&&&\\
&&& \lstick{\ket{x_{0}}_{3}} & \qw & \ctrl{2} & \targ    & \qswap  & \qw & \qw & \qw & \qw & \qw & \qw & \qw & \qw & \rstick{\ket{x''_{1,0}}_{3}}\\
&&&&&&&&&&&&&&&&&&&\\
&&& \lstick{\ket{0}_{1}}     & \qw & \targ    & \ctrl{3} & \ctrl{3}& \qw & \qw & \qw & \rstick{\ket{x_{0}}_{1}}\\
&&& \lstick{\ket{0}_{2}}     & \qw & \qw      & \qw      & \qswap  & \qw & \qw & \qw & \qw & \qw & \qw & \qw & \qw & \rstick{\ket{x'_{0,0}}_{2}}\\
&&&&&&&&&&&&&&&&&&&\\
&&& \lstick{\ket{0}_{3}}     & \qw & \qw      & \targ    & \qswap  & \qw & \qw & \qw & \qw & \qw & \qw & \qw & \qw & \rstick{\ket{x''_{0,0}}_{3}}\\
}}}
\end{minipage}
\end{center}
\caption{\label{encoder4}4-qubits Quantum Booth Encoder.}
\end{figure}
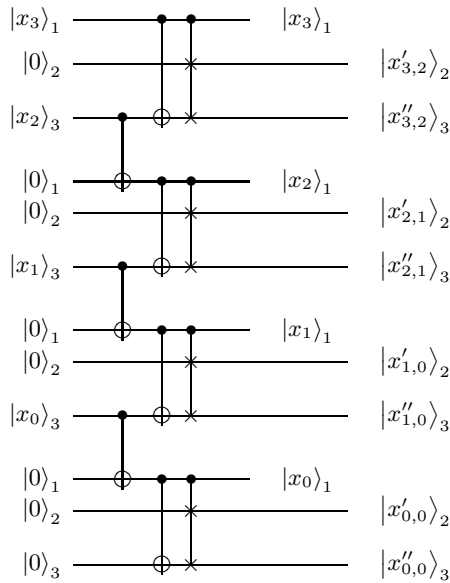

If we would like to encode a 4-qubits binary chain, for instance,
we should built the Booth Encoder circuit ($\mathcal{BE}$) shown
in Fig.~\ref{encoder4} and, for reversibility purposes, apply its
inverse ($\mathcal{BE}^{-1}$), which is the same circuit but using
the outputs as inputs, and viceversa

Now, for our 4-qubits example, the QBM architecture can be
described as follows. The quantum circuit shown in
Fig.~\ref{encoder4} encodes the binary transitions ($i+1,i$) and
controls, by Toffoli gates, what will be stored at the quantum
registers, where the partial products have been loaded. Hence,
depending on the transitions in the quantum binary chain, we will
perform one out of the three following operations over the
multiplicand:

\begin{itemize}
\item
If we have no transition, $|00\rangle$, then load 0's in the
partial product register.
\item
If we have up-transitions, $|01\rangle$, then load the
multiplicand in the partial product register.
\item
If we have down-transitons, $|10\rangle$, then load the
multiplicand 1's-complement representation in the partial product
register.
\end{itemize}

Finally, we have to sum all the partial products that we have
generated. If the multiplicand and the multiplier have $n$ qubits,
each partial product will have $2n$ qubits and they will be
shifted, from the encoded multiplier Less Significant Bit (LSB) to
the Most Significant Bit (MSB), depending on the ordinal position
of the multipliers bits. These displacements are implemented
within the quantum $2n$-registers hardware.

After the $\cal{O}(\log$\,n$)$ sums have been performed we still
have to do something more because we have used 1's-complement
representation instead of 2's-complement representation. Indeed,
we have saved the ``carry'' bit bearing in mind that
2's-complement representation is equal to 1's-complement
representation plus one. Therefore, we have to add, to the whole
partial products sum, another register where the 1's have been
stored in their ordinal proper way, as one can see in the example
displayed in Fig.~\ref{multiplicador}. Therefore, the final output
is the desired input product represented in 2's-complement.

\section{Conclusions}

\begin{figure} [htbp]
\vspace*{13pt}
\begin{center}
\begin{minipage}[l]{13cm}
\flushleft
\newenvironment{vardesc}[1]{%
\settowidth{\parindent}{#1\ } \makebox[3pt][l]{#1\ }}{}
\begin{vardesc}
{\scriptsize{\Qcircuit @C=.5em @R=.7em { &\lstick{\ket{x_{3}}}&
\qw & \multigate{7}{\mathcal{BE}} & \qw &\qw & \ctrl{22} &
\ctrl{1} & \ctrlo{1} & \qw & \qw & \qw & \qw & \qw & \qw & \qw &
\qw & \qw & \qw & \qw & \qw & \qw & \qw & \qw & \ctrl{11} &
\qw & \qw & \qw & \multigate{7}{\mathcal{BE}^{-1}} & \qw & \rstick{\ket{x_{3}}}\\
&\lstick{\ket{x_{2}}}& \ctrl{1} & \ghost{\mathcal{BE}} & \qw
&\ctrl {21} & \qw & \ctrlo{22}& \ctrlo{22}& \qw & \qw & \qw & \qw
& \qw & \qw & \qw & \qw & \qw & \qw & \qw & \qw & \qw & \qw & \qw
& \qw & \qw
& \qw & \qw & \ghost{\mathcal{BE}^{-1}} & \qw & \rstick{\ket{x_{2}}}\\
&&& \ghost{\mathcal{BE}} &\qw & \qw & \qw & \qw & \qw & \qw &
\ctrl {25} & \ctrl{1} & \ctrlo{1} & \qw & \qw & \qw & \qw & \qw &
\qw & \qw
& \qw & \qw & \qw & \qw & \qw & \ctrl{9} & \qw & \qw & \ghost{\mathcal{BE}^{-1}}\\
&\lstick{\ket{x_{1}}}&\ctrl{1}& \ghost{\mathcal{BE}} & \qw &\qw &
\qw & \qw & \qw & \ctrl{24} & \qw & \ctrlo{25}& \ctrlo{25} & \qw &
\qw & \qw & \qw & \qw & \qw & \qw & \qw & \qw & \qw & \qw & \qw &
\qw & \qw
& \qw & \ghost{\mathcal{BE}^{-1}} & \qw & \rstick{\ket{x_{1}}}\\
&&& \ghost{\mathcal{BE}} &\qw & \qw & \qw & \qw & \qw & \qw & \qw
& \qw & \qw & \qw & \ctrl{28} & \ctrl{1} & \ctrlo{1} & \qw & \qw &
\qw
& \qw & \qw & \qw & \qw & \qw & \qw & \ctrl{7} & \qw & \ghost{\mathcal{BE}^{-1}}\\
&\lstick{\ket{x_{0}}}&\ctrl{1}& \ghost{\mathcal{BE}} & \qw &\qw &
\qw & \qw & \qw & \qw & \qw & \qw & \qw & \ctrl{27} & \qw &
\ctrlo{28}& \ctrlo{28} & \qw & \qw & \qw & \qw & \qw & \qw & \qw &
\qw & \qw &
\qw & \qw & \ghost{\mathcal{BE}^{-1}} & \qw & \rstick{\ket{x_{0}}}\\
&&& \ghost{\mathcal{BE}} &\qw & \qw & \qw & \qw & \qw & \qw & \qw
& \qw & \qw & \qw & \qw & \qw & \qw & \qw & \ctrl{31} & \ctrl{1} &
\ctrlo{1} & \qw & \qw & \qw & \qw & \qw & \qw & \ctrl{5} & \ghost{\mathcal{BE}^{-1}}\\
&\lstick{\ket{\mathbf{0}}}& \qw & \ghost{\mathcal{BE}} &\qw & \qw
& \qw & \qw & \qw & \qw & \qw & \qw & \qw & \qw & \qw & \qw & \qw
& \ctrl{30} & \qw & \ctrlo{31}& \ctrlo{31} & \qw & \qw & \qw & \qw
& \qw & \qw &
\qw & \ghost{\mathcal{BE}^{-1}} & \qw & \rstick{\ket{\mathbf{0}}}\\
&&&&&&&&&&&&&&&&&&&\\
&&&&&&&&&&&&&&&&&&&\\
&&&&&&&&&&&&&&&&&&&&&&\lstick{\ket{\hat{0}_{(4)}}}&{/}\qw & \qw &
\qw
& \qw & \qw & \qw & \multigate{1}{\mathcal{C}_{2}}\\
&&&&&&&&&&&&&&&&&&&&&&\lstick{\ket{\hat{0}_{(4)}}}&{/}\qw & \targ
&
\targ & \targ & \targ & \qw & \ghost{\mathcal{C}_{2}}\\
&&&&&&&&&&&&&&&&&&&&&&&&{(3)}&{(2)}&{(1)}&{(0)}\\
&&&&&&&&&&&&&&&&&&&\\
&&&&&&&&&&&&&&&&&&&\\
&&&&&&&&&&&&&&&&&&&\\
&&& \lstick{\ket{y_{3}}} & \qw & \control\qw & \control\qw & \qw &
\qw & \control\qw & \control\qw & \qw & \qw & \control\qw &
\control
\qw & \qw & \qw & \control\qw & \control\qw & \qw & \qw & \qw\\
&&&&&&&&&&&&&&&&&&&\\
&&& \lstick{\ket{\hat{y}_{(4)}}} &{/}\qw & \qw & \qw & \qw &
\control \qw & \qw & \qw & \qw & \control \qw & \qw & \qw & \qw &
\control \qw
& \qw & \qw & \qw & \control \qw & \qw\\
&&&&&&&&&&&&&&&&&&&\\
&&&&&&&&&&&&&&&&&&&\\
&&&&&&&&&&&&&&&&&&&\\
&&& \lstick{\ket{0}} & \qw & \targ & \targ & \targ & \qw & \qw &
\qw & \qw & \qw & \qw & \qw & \qw & \qw & \qw & \qw & \qw & \qw &
\multigate{2}{\mathcal{S}_{3}} & \qw & \qw & \multigate{7}{\Sigma}\\
&&& \lstick{\ket{\hat{y}_{(4)}}} &{/}\qw & \qw & \qw & \targ &
\targ & \qw & \qw & \qw & \qw & \qw & \qw & \qw & \qw & \qw & \qw
& \qw &
\qw & \ghost{\mathcal{S}_{3}} & \qw & {/}\qw & \ghost{\Sigma}\\
&&& \lstick{\ket{\hat{0}_{(3)}}} &{/}\qw & \qw & \qw & \qw & \qw &
\qw & \qw & \qw & \qw & \qw & \qw & \qw & \qw & \qw & \qw & \qw &
\qw
& \ghost{\mathcal{S}_{3}} & \qw & {/}\qw & \ghost{\Sigma}\\
&&&&&&&&&&&&&&&&&&&\\
&&&&&&&&&&&&&&&&&&&\\
&&& \lstick{\ket{\hat{0}_{(2)}}} &{/}\qw & \qw & \qw & \qw & \qw &
\targ & \targ & \targ & \qw & \qw & \qw & \qw & \qw & \qw & \qw &
\qw & \qw & \multigate{2}{\mathcal{S}_{2}} & \qw & {/}\qw & \ghost
{\Sigma} & {/}\qw & \gate{\mathcal{S}_{2}+\mathcal{S}_{3}} &
\multigate{7}{\Sigma}\\
&&& \lstick{\ket{\hat{y}_{(4)}}} &{/}\qw & \qw & \qw & \qw & \qw &
\qw & \qw & \targ & \targ & \qw & \qw & \qw & \qw & \qw & \qw &
\qw &
\qw & \ghost{\mathcal{S}_{2}} & \qw & {/}\qw & \ghost{\Sigma}\\
&&& \lstick{\ket{\hat{0}_{(2)}}} &{/}\qw & \qw & \qw & \qw & \qw &
\qw & \qw & \qw & \qw & \qw & \qw & \qw & \qw & \qw & \qw & \qw &
\qw
& \ghost{\mathcal{S}_{2}} & \qw & {/}\qw & \ghost{\Sigma}\\
&&&&&&&&&&&&&&&&&&&&&&&&&&&& {=} & {\mathcal{{S}}_{Tot.}}\\
&&&&&&&&&&&&&&&&&&&&&&&&&&&&&\\
&&& \lstick{\ket{\hat{0}_{(3)}}} &{/}\qw & \qw & \qw & \qw & \qw &
\qw & \qw & \qw & \qw & \targ & \targ & \targ & \qw & \qw & \qw &
\qw & \qw & \multigate{2}{\mathcal{S}_{1}} & \qw & {/}\qw &
\multigate{6}
{\Sigma}\\
&&& \lstick{\ket{\hat{y}_{(4)}}} &{/}\qw & \qw & \qw & \qw & \qw &
\qw & \qw & \qw & \qw & \qw & \qw & \targ & \targ & \qw & \qw &
\qw &
\qw & \ghost{\mathcal{S}_{1}} & \qw & {/}\qw & \ghost{\Sigma}\\
&&& \lstick{\ket{0}} & \qw & \qw & \qw & \qw & \qw & \qw & \qw &
\qw & \qw & \qw & \qw & \qw & \qw & \qw & \qw & \qw & \qw & \ghost
{\mathcal{S}_{1}} & \qw & \qw & \ghost{\Sigma} & {/}\qw & \gate
{\mathcal{S}_{0}+\mathcal{S}_{1}} & \ghost{{\Sigma}}\\
&&&&&&&&&&&&&&&&&&&\\
&&&&&&&&&&&&&&&&&&&\\
&&& \lstick{\ket{\hat{0}_{(4)}}} &{/}\qw & \qw & \qw & \qw & \qw &
\qw & \qw & \qw & \qw & \qw & \qw & \qw & \qw & \targ & \targ &
\targ
& \qw & \multigate{1}{\mathcal{S}_{0}} & \qw & {/}\qw & \ghost{\Sigma}\\
&&& \lstick{\ket{\hat{y}_{(4)}}} &{/}\qw & \qw & \qw & \qw & \qw &
\qw & \qw & \qw & \qw & \qw & \qw & \qw & \qw & \qw & \qw & \targ
&
\targ & \ghost{\mathcal{S}_{0}} & \qw & {/}\qw & \ghost{\Sigma}\\
}}}
\end{vardesc}
\end{minipage}
\end{center}
\vspace*{13pt} \caption{\label{multiplicador}4-qubits Quantum
Booth's array multiplier circuit}
\end{figure}
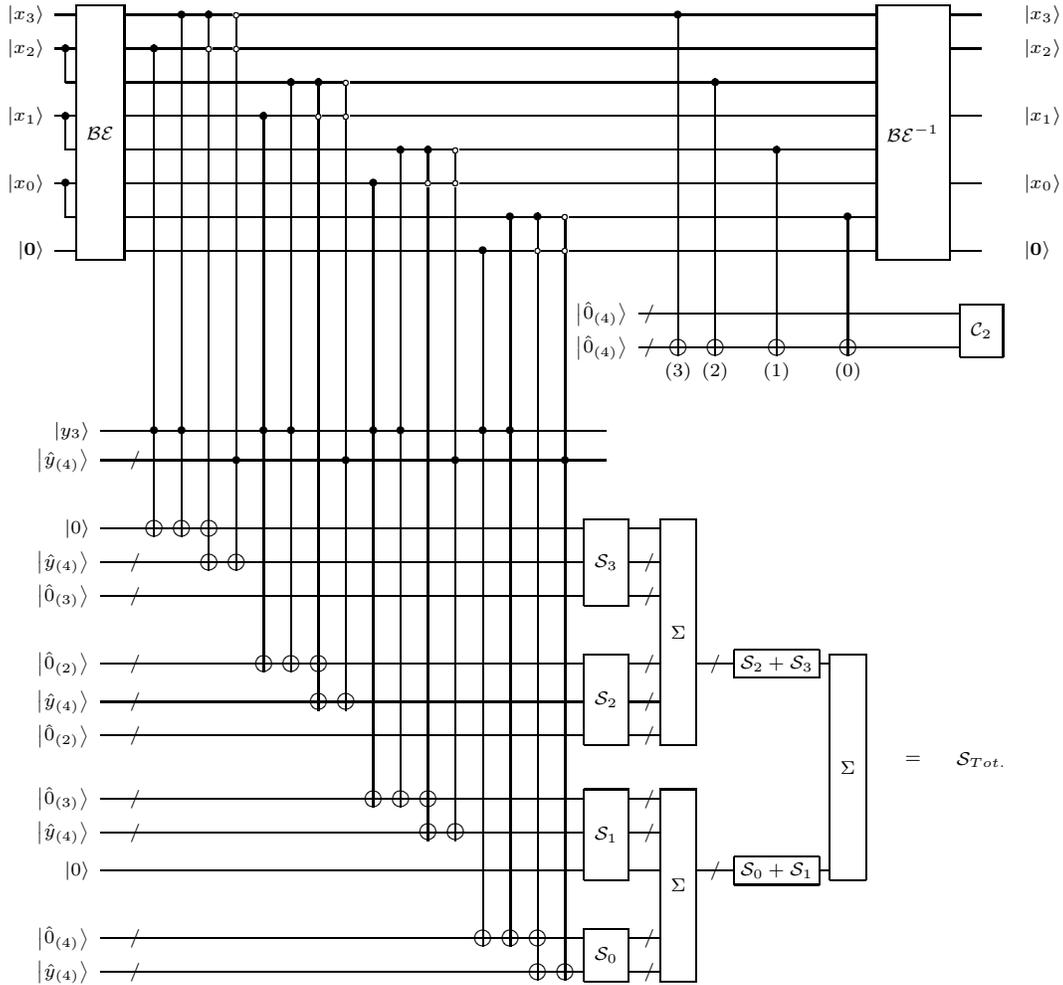

In this work a new Quantum Booth Algorithm is presented. Note that
QBM has a very regular structure. It can be easily implemented in
order to build bigger circuits for bigger inputs than the one
shown in Fig.~\ref{multiplicador}. Its delay is $\cal{O}((\log
$\,n$)^{2})$ due to the adders structure are $\cal{O}(\log$\,n$)$
and each adder has a logarithmic deph. One could reduces it by
means of a CSA scheme \cite{Sunar} but it would made to lose the
circuit regular structure, which is an undesireble compensation.
Thus, the bottleneck are the adders and our future researching
efforts will go in the direction of clarifying whether
$\cal{O}(\log$\,n$)$ is a fundamental bound, from the quantum
physics point of view, or not. Note that adders information
processing \cite{logarithmic-carry-add} is classical due to that
the input information is classical; there is nor entanglement
neither quantum superposition. Therefore, the Quantum Booth
Algorithm architecture is also a classical reversible one because
the quantum gates used in the encoding and control stages are also
classical reversible ones (TOFFOLI, SWAP, CNOT). No Hadamards are
presented.

The main target of this paper was building a quantum algorithm for
the multiplication task using the fundamental features that
quantum computation has: entanglement and superposition of states.
But, when we tried this way, we realized that we did not need
quantum entanglement for building our quantum multiplying
algorithm. The reason was that we were \emph{rewriting} a
classical irreversible algorithm by means of classical reversible
gates (excluding Hadamard ones). Indeed, we believe that
for obtaining a real improvement, it is necessary to reconsider the way in which  the Quantum Arithmetic Logic
Unit was designed, in order to perform the quantum arithmetic operations taking into account, as essential ingredients, quantum entanglement and quantum
superposition. Work in this direction is in progress.

\begin{acknowledgments}
J.J. \'Alvarez-S\'anchez
thanks Profs.  C. G\'omez, G. Brassard and J. M. Fern\'andez for
valuable discussions and advice. This work has been partially
supported by Spanish Ministerio de Educaci\'on y Ciencia (Project
MTM2005-09183) and Junta de Castilla y Le\'on (Excellence Project
VA013C05). The Q-circuit package was used to produce the quantum
circuits included in this paper.
\end{acknowledgments}

\end{document}